\documentclass[prl,twocolumn,showpacs]{revtex4}
\usepackage{times}
\usepackage{graphicx}

\def\kbar{\protect\@kbar}
\def\@kbar{%
\relax \bgroup
\def\@tempa{\hbox{\raise.73\ht0
\hbox to0pt{\kern.25\wd0\vrule width.5\wd0
height.1pt depth.1pt\hss}\box0}}%
\mathchoice{\setbox0\hbox{$\displaystyle k$}\@tempa}%
{\setbox0\hbox{$\textstyle k$}\@tempa}%
{\setbox0\hbox{$\scriptstyle k$}\@tempa}%
{\setbox0\hbox{$\scriptscriptstyle k$}\@tempa}%
\egroup}

\begin{document}

\title{Signatures of quantum stability in a classically chaotic system}
\author{S. Schlunk,$^{1}$ M.B. d'Arcy$,^{1}$ S.A.
Gardiner,$^{1}$ D. Cassettari,$^{1}$ R.M. Godun,$^{1}$ and G.S.
Summy$^{1,2}$}

\affiliation{$^{1}$Clarendon Laboratory, Department of Physics,
University of Oxford, Parks Road, Oxford, OX1 3PU, United Kingdom
\\ $^2$Department of Physics, Oklahoma State University,
Stillwater, Oklahoma, 74078-3072}

\date{\today}

\begin{abstract}
We experimentally and numerically investigate the quantum
accelerator mode dynamics of an atom optical realization of the
quantum $\delta$-kicked accelerator, whose classical dynamics are
chaotic. Using a Ramsey-type experiment, we observe interference,
demonstrating that quantum accelerator modes are formed
coherently. We construct a link between the behavior of the
evolution's fidelity and the phase space structure of a recently
proposed pseudoclassical map, and thus account for the observed
interference visibilities.
\end{abstract}


\maketitle

The relationship between the behaviour of classical and quantum
systems, and how macroscopic classical phenomena originate in the
quantum regime, remain subjects of dispute \cite{zurek91}. The
issues involved are particularly marked for quantum versions of
classically chaotic systems \cite{haake2001}. Experimental
investigations of such systems began with studies of
microwave-driven hydrogen \cite{koch95}; subsequent work has also
centred on microwave cavities \cite{sridhar92}, mesoscopic
solid-state systems \cite{wilkinson96}, and atom optics
\cite{moore95}, the approach we adopt. In this Letter we consider
the quantum $\delta$-kicked accelerator
\cite{oberthaler99,godun2000,darcy2001a}, a $\delta$-kicked rotor
with an additional static linear potential. The $\delta$-kicked
rotor is one of the most extensively investigated systems in
chaotic dynamics \cite{lichtenberg92}, and is equivalent to a free
particle subjected periodically to instantaneous momentum kicks
from a sinusoidal potential. Quantum mechanically, the effect of
these kicks is to diffract the particles' constituent de Broglie
waves into a series of discrete momentum states. In the
$\delta$-kicked accelerator, the linear potential modifies the
chaotic classical dynamics only slightly, yet can radically change
the quantum behaviour. The phases accumulated between consecutive
kicks by the momentum states are altered, leading to the creation
of quantum accelerator modes
\cite{oberthaler99,godun2000,darcy2001a}. We realize quantum
$\delta$-kicked accelerator dynamics in laser-cooled cesium atoms
by the application of short pulses of a vertical standing wave of
off-resonant laser light, which constitutes a sinusoidal
potential; gravity provides the linear potential. Quantum
accelerator modes, absent in the analogous classical dynamics, are
observed and are characterized by a linear (with kick number)
momentum transfer to a substantial fraction ($\sim 20\%$) of the
atoms. If coherent, this efficient momentum transfer promises
applications in atom interferometry \cite{berman97}. In this
Letter we use a Ramsey-type interference experiment
\cite{ramsey86} to show that quantum accelerator modes do preserve
coherence. We then relate the Ramsey fringe contrast to the
fidelity $f$ \cite{casati2002}; by a numerical analysis, we link
the behaviour of $f$ to the phase space structure of the
$\delta$-kicked accelerator in a pseudoclassical limit recently
proposed by Fishman {\em et al.} \cite{fishman2002}. Finally we
explain differences in the observed fringe visibilities by
examining the effect of the experimental range of kicking
strengths.

In our interference experiment, the atoms undergo $\delta$-kicked
accelerator dynamics, between the application of two $\pi/2$
microwave pulses that couple two atomic hyperfine levels. In the
absence of coherence-destroying spontaneous emission, the contrast
of any interference fringes is related to the overlap of two
initially identical motional states evolved under the influence of
slightly different Hamiltonians \cite{gardiner97}, i.e. the
fidelity. It can therefore yield information on the sensitivity of
the atoms' evolution to variations in the kicking strength. Strong
sensitivity can be considered a quantum signature of chaos,
particularly in the semiclassical limit ($\hbar \rightarrow 0$),
hence the use by Peres \cite{peres93} of $f$ as a measure of
quantum stability.

\begin{figure}
\begin{center}
\includegraphics[width=3.3in]{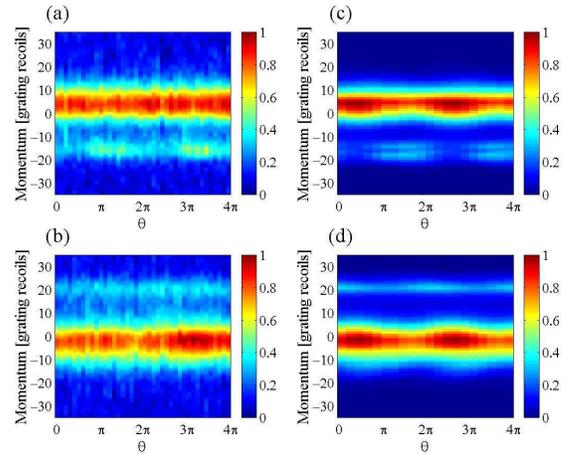}
\end{center}
\caption{Experimentally measured momentum distributions as the
microwave phase difference $\theta$ is varied in a $\pi/2$ ---
$20$ kick
--- $\pi/2$ sequence where $\delta_{L}^{b}\sim 35$\thinspace GHz,
with (a) $T=60.5\mu$s (accelerator mode at $-17 \hbar G$), and (b)
$T=74.5\mu$s (accelerator mode at $20 \hbar G$). The corresponding
numerically generated momentum distributions, where
$\delta_{L}^{b}=35$\thinspace GHz, are shown in (c) and (d).
Population has been arbitrarily normalized to maximum value $=1$.}
\label{Fig:experiment theory}
\end{figure}

After magneto-optic trapping and molasses cooling to $5\mu$K, we
prepare around $10^{6}$ freely falling cesium atoms in the $F = 3,
m_F = 0$  hyperfine level (denoted $|a\rangle$) of the
$6^{2}$S$_{1/2}$ ground state \cite{amcomment}. The first $\pi/2$
microwave pulse creates an equal superposition of the atoms'
internal states, i.e., $ |a\rangle \rightarrow (|a\rangle -
ie^{i\theta}|b\rangle)/\sqrt{2}$, where $|b\rangle$ denotes the $F
= 4, m_F = 0$ level. The phase $\theta$ of this pulse can be
changed with respect to that of the second $\pi/2$ pulse, applied
following 20 equally spaced 500\thinspace ns pulses from a
standing wave of light. This is formed by retroreflection of a
Ti:sapphire laser beam; its maximum intensity is $\sim 1\times
10^{8}$\thinspace mW/cm$^{2}$ \cite{darcy2001a}, and the light is
red-detuned by $45$ and $35$\thinspace GHz from the D1 transition
for atoms in states $|a\rangle$ and $|b\rangle$, respectively.
After the second $\pi/2$ microwave pulse, we measure the momentum
distribution in state $|b\rangle$ by a time-of-flight method. For
more details of our experimental setup see Refs.\
\cite{godun2000,darcy2001a}. Measurement of a periodic variation
with $\theta$ in the accelerator mode population in state
$|b\rangle$, i.e.\ interference, directly implies coherent
evolution.

In the limit of large detuning, the Hamiltonian is
\begin{equation}
\hat{H} = \hat{H}_{a}|a\rangle\langle a|
+\hat{H}_{b}|b\rangle\langle b| +
\frac{\hbar\omega_{ab}}{2}(|b\rangle\langle b|-|a\rangle\langle
a|),
\end{equation}
where $\hbar\omega_{ab}$ is the energy gap between $|a\rangle$ and
$|b\rangle$, and
\begin{equation}
    \hat{H}_{\sigma} = \frac{\hat{p}^{2}}{2m} + mg\hat{z} -
    \frac{\hbar\Omega^{2}t_{p}}
    {8\delta_{L}^{\sigma}} [1 + \cos(G\hat{z})]\sum_{n}\delta(t-nT)
\end{equation}
is the quantum $\delta$-kicked accelerator Hamiltonian, acting on
atoms in internal state $|\sigma\rangle \in \{|a\rangle,
|b\rangle\}$. Here $\hat{z}$ is the vertical position, $\hat{p}$
the $z$-direction momentum, $m$ the particle mass, $g$ the
gravitational acceleration, $t$ the time, $T$ the pulsing period,
$\Omega$ the Rabi frequency, $t_{p}$ the pulse duration,
$\delta_{L}^{\sigma}$ the detuning from the D1 transition for the
state $|\sigma\rangle$, and $G = 4\pi / \lambda$, where
$\lambda=894$\thinspace nm is the laser wavelength, and $\hbar G$
is a grating recoil (the momentum separation of adjacent
diffracted states). We denote the amplitude of the phase
modulation to atoms in state $|b\rangle$ that results from
application of the standing wave as
$\phi_{d}=\Omega^{2}t_{p}/8\delta_{L}^{b}$. The experimental mean
value of $\phi_{d}$ is $0.8\pi$, and, due to the different
detuning, that of the corresponding quantity for atoms in state
$|a\rangle$ is $\phi_{d}^{a}=
\phi_{d}\delta_{L}^{b}/\delta_{L}^{a}=0.6\pi$. We thus have
effectively two different Hamiltonians, applied to the same
initial motional state. The pulse train leads to the creation of a
quantum accelerator mode, the momentum of which is the same for
the two internal states \cite{godun2000}. We consider pulse
periods $T=60.5\mu$s and $74.5\mu$s, close to $T_{1/2} = 2\pi
m/\hbar G^{2}= 66.7\mu$s, which corresponds to the lowest
second-order quantum resonance in the $\delta$-kicked rotor
\cite{darcy2001a,oskay2000}. For these $T$, well-populated
accelerator modes involving substantial momentum transfer are
created \cite{oberthaler99,godun2000,darcy2001a}.

Figure \ref{Fig:experiment theory}(a) shows the measured final
momentum distributions of $|b\rangle$ atoms, for $T=60.5\mu$s. We
see a period-$2\pi$ variation with $\theta$ in the accelerator
mode population (at around $-17 \hbar G$), the visibility $V$ of
which is $(21 \pm 2)\%$ \cite{visibility}. We observe similar
fringes for a range of detunings
($\delta_{L}^{b}=20$--40\thinspace GHz) and total number of kicks
$N=10$--30. However, depending on their exact values, $V$ can vary
between $10\%$ and $40\%$. This periodic variation of the
population demonstrates interference, and hence that the
accelerator mode transfers momentum coherently. At $T = 74.5\mu$s
[Fig.\ \ref{Fig:experiment theory}(b)], however, fringes in the
accelerator mode (at around $20 \hbar G$) are practically
invisible, despite the expected coherent nature of the momentum
transfer. In Figs.\ \ref{Fig:experiment theory}(c) and
\ref{Fig:experiment theory}(d) diffraction-based numerical
simulations \cite{oberthaler99,godun2000,darcy2001a},
incorporating the experimental range of $\phi_{d}$ ($0.3\pi$ to
$1.2\pi$), also show this difference in the fringe visibility for
the two values of $T$. The range of $\phi_{d}$ is due to the
Gaussian profile of the standing wave intensity (FWHM $\sim
1$\thinspace mm) and the spatial extent of the atomic cloud
(Gaussian density distribution, FWHM $\sim 1$\thinspace mm)
\cite{darcy2001a}. As we optimized the overlap of the laser beams
with the atomic cloud, the intensity and density maxima can be
assumed to be coincident. The calculated visibility is then $25\%$
for $T=60.5\mu$s [Fig.\ \ref{Fig:experiment theory}(c)] but only
$8\%$ for $T=74.5\mu$s [Fig.\ \ref{Fig:experiment theory}(d)].

In order to explain these surprising observations, we introduce
the Floquet operator $\hat{F}_{b}(\phi_{d})$. This describes the
effect of one kick and the subsequent free evolution on the
motional state of atoms in state $|b\rangle$, where
\begin{equation}\label{Eq:Floquetoperator1}
    \hat{F}_{b}(\phi_{d}) = \exp(-i[\gamma\hat{\chi}+
    \hat{\rho}^{2}/2]/\kbar)\exp (i \phi_d[1+ \cos
    \hat{\chi}]).
\end{equation}
We define $\hat{F}_{a}(\phi_{d})$ analogously for state
$|a\rangle$, with $\phi_{d}$ replaced by $\phi_{d}^{a}$
\cite{phinote}. As in Ref.\ \cite{darcy2001a}, we use scaled
position and momentum variables $\chi = Gz$ and $\rho = GTp/m$,
while $\gamma=gGT^{2}$ describes the effect of gravity, and $\kbar
=\hbar G^2 T/m = -i[\hat{\chi},\hat{\rho}]$ is an effective scaled
Planck constant. After $N$ pulses an initial plane wave
$|q\rangle$ of wavenumber $q$ evolves to
$\hat{F}_{\sigma}(\phi_{d})^{N}|q\rangle = e^{i\phi
N}|\psi_{\sigma}^{q}(\phi_{d})\rangle$, where $\phi= \phi_{d}$ or
$\phi_{d}^{a}$, as appropriate. Regarding the initial motional
state as an incoherent superposition of $|q\rangle$, the momentum
distribution in state $|b\rangle$ for a given $\phi_{d}$ after the
$\pi/2$ --- $N$ kick
--- $\pi/2$ sequence, is
\begin{eqnarray}\label{eq:simons calculation}
    \nonumber
    P_{b}(\phi_{d},p) &=&
    \frac{1}{4}
    \int dq
    C(q)\left[|{\psi}_{a}^{q}(\phi_{d},p)|^2
    +
    |\psi_{b}^{q}(\phi_{d},p)|^2\right]\\
     && + \frac{1}{2}\left |
    \int dq C(q) \psi_{a}^{q}(\phi_{d},p)^{*}
    \psi_{b}^{q}(\phi_{d},p)
\right |
    \nonumber \\ && \times
    \cos(\phi_{I}(\phi_{d},p)+N\delta\phi_d +\theta),
\end{eqnarray}
where $\delta\phi_{d} = \phi_{d}-\phi_{d}^{a}$,
$\psi_{\sigma}^{q}(\phi_{d},p)=\langle
p|\psi_{\sigma}^{q}(\phi_{d})\rangle$, and $\phi_I$ is the phase
of the interference term, i.e., $\int dq C(q) {\psi}_{a}^{q*}
\psi_{b}^{q}=|\int dq C(q) {\psi}_{a}^{q*} \psi_{b}^{q}
|e^{i\phi_I}$. The weighting $C(q)$ describes the initial Gaussian
momentum distribution (FWHM $=6$ grating recoils $\hbar G$). The
third (interference) term  in Eq.\ (\ref{eq:simons calculation})
is responsible for the appearance of fringes in the accelerated
$|b\rangle$ population. We denote the amplitude of the modulation
in $P_{b}$ by $A(\phi_{d},p)/2=|\int dq C(q)
{\psi}_{a}^{q}(\phi_{d},p)^{*} \psi_{b}^{q}(\phi_{d},p)|/2$, where
$\int dp A(\phi_{d},p)^{2}=f(\phi_{d})$ is the fidelity for a
given $\phi_{d}$.

\begin{figure}
\begin{center}
\includegraphics[width=3.3in]{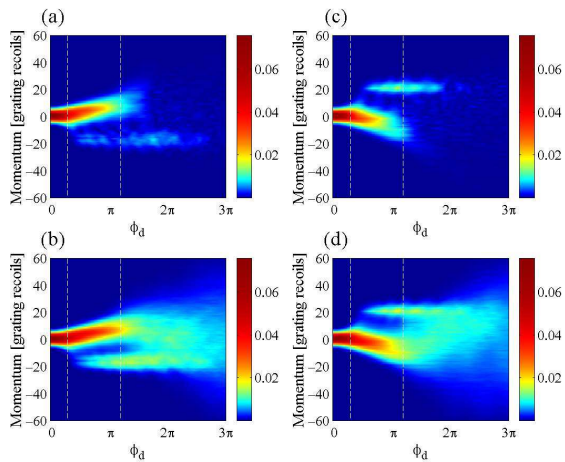}
\end{center}
\caption{Plots of $A/2$ against $\phi_{d}$ with $N=20$, for (a) $T
= 60.5\mu$s and (c) $T = 74.5\mu$s. Plots of the non-interfering
population $\int dq C(q)(|\psi_{a}^{q}|^{2}+|\psi_{b}^{q}|^{2})/4$
for (b) $T = 60.5\mu$s and (d) $T = 74.5\mu$s. Dashes mark the
boundaries of the experimental range. Accelerator modes exist at
$-17 \hbar G$ for (a) and (c), and at $20 \hbar G$ for (b) and
(d).} \label{Fig:psi1 psi2}
\end{figure}

We have calculated the individual terms of $P_b$ numerically for a
wide range of $\phi_d$, obtaining as a consequence an important
result linking the pseudoclassical analysis of Fishman {\it et
al}.\ \cite{fishman2002} with the quantum stability measure of
Peres \cite{peres93}. Comparison of Figs.\ \ref{Fig:psi1 psi2}(a)
and \ref{Fig:psi1 psi2}(c) with Fig.\ \ref{Fig:experiment theory}
shows that the region in momentum space corresponding to an
accelerator mode is also a region of high $A$. This remains high
up to large values of $\phi_{d}$ \cite{stability}, continuing
beyond the point at which it has decayed to nearly zero in other
regions of momentum space. As $f = \int dp A^{2}$, its large value
when determined by integrating over the momenta populated by atoms
in the quantum accelerator mode implies that these atoms inhabit a
stable region of quantum state space. Note that small $A$ does not
necessarily imply low atomic population, as can be seen in the
plots of the non-interfering population $\int dq
C(q)(|\psi_{a}^{q}|^{2}+|\psi_{b}^{q}|^{2})/4$ in Figs.\
\ref{Fig:psi1 psi2}(b) and \ref{Fig:psi1 psi2}(d). Contrasting
Fig.\ \ref{Fig:psi1 psi2}(a) with Fig.\ \ref{Fig:psi1 psi2}(c), we
see that this large value of $A$ extends over a significantly
wider range of $\phi_{d}$ for $T=60.5\mu$s than for $T=74.5\mu$s.
Hence, we can interpret the accelerator mode at $T=60.5\mu$s as
being more robust to variations in $\phi_{d}$, i.e. more {\it
stable}, compared with that at $T=74.5\mu$s. However given our
comparatively narrow experimental range of $\phi_{d}$, this does
not explain the difference in fringe visibilities seen in Fig.\
\ref{Fig:experiment theory}.

\begin{figure}
\begin{center}
\includegraphics[width=3.3in]{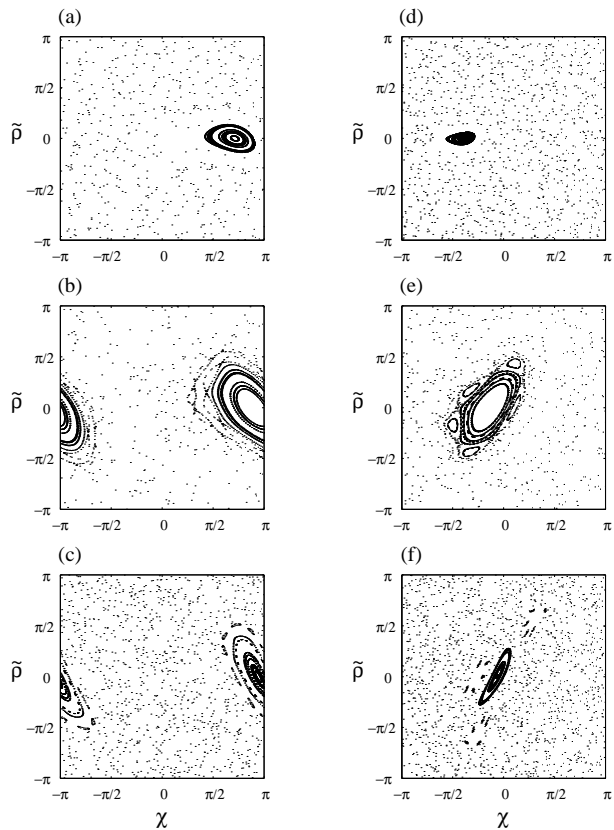}
\end{center}
\caption{Stroboscopic pseudoclassical Poincar\'{e} sections
determined by Eqs.\ (\ref{Eq:rhomap}) and (\ref{Eq:chimap}) for
$T=60.5\mu$s ($\Rightarrow \epsilon = -0.58$), and for (a)
$\phi_d=0.35\pi$, (b) the average experimental value $0.8\pi$, and
(c) $1.5\pi$. Figures (d), (e), and (f) show corresponding plots
for $T=74.5\mu$s ($\Rightarrow \epsilon = 0.73$). Displayed units
are dimensionless.} \label{Fig:phasespace}
\end{figure}

The appearance of quantum accelerator modes in the $\delta$-kicked
accelerator is explained in the analysis of Fishman {\it et al}.\
\cite{fishman2002} in terms of islands of stability centred on
stable fixed points in the phase space generated by the map
\cite{mapnote}:
\begin{eqnarray}
\tilde{\rho}_{n+1} &=& \tilde{\rho}_{n} - \tilde{k}\sin(\chi_{n})
-\mbox{sign}(\epsilon)\gamma, \label{Eq:rhomap}\\ \chi_{n+1} &=&
\chi_{n} + \mbox{sign}(\epsilon)\tilde{\rho}_{n+1},
\label{Eq:chimap}
\end{eqnarray}
where the population of a mode is proportional to the size of the
corresponding island. This is a {\it pseudoclassical\/}
[$\epsilon=(\kbar-2\pi)\rightarrow 0$] rather than {\it
semiclassical\/} ($\kbar \rightarrow 0$) limit of the quantum
dynamics characterized by the Floquet operator of Eq.\
(\ref{Eq:Floquetoperator1}). We have introduced $\tilde{\rho}=\rho
\epsilon/\kbar$ (in an accelerating frame \cite{fishman2002}) and
$\tilde{k}=\phi_{d}|\epsilon|$. Classically, the system is
globally chaotic for our parameter regime.
Figure~\ref{Fig:phasespace} shows the pseudoclassical phase spaces
generated by iteration of Eqs.\ (\ref{Eq:rhomap}) and
(\ref{Eq:chimap}) for the experimentally investigated values of
$\epsilon = 2\pi(T/T_{1/2}-1)$, and a range of $\phi_{d}$. When
$\phi_{d}=0.3\pi$ [Figs.\ \ref{Fig:phasespace}(a) and
\ref{Fig:phasespace}(d)], the island (based around a stable fixed
point in phase space) is substantially smaller for $T=74.5\mu$s
than for $T=60.5\mu$s. For the average experimental value of
$\phi_{d}$ ($0.8\pi$) [Figs.\ \ref{Fig:phasespace}(b) and
\ref{Fig:phasespace}(e)], the islands have both grown to be about
the same size.  For $\phi_{d}=1.5\pi$ [Figs.\
\ref{Fig:phasespace}(c) and \ref{Fig:phasespace}(f)], the island
has shrunk dramatically in the case of $T=74.5\mu$s, while at
$T=60.5\mu$s the island has also shrunk, but not to the same
extent. We therefore conclude that the stable island representing
the accelerator mode is much more robust to perturbations in the
kicking strength for $T=60.5\mu$s than for $T=74.5\mu$s. The fact
that $A$ (and therefore $f$) remains large at the accelerator mode
momentum for a significantly broader range of $\phi_{d}$ when
$T=60.5\mu$s than for when $T=74.5\mu$s, as shown in Fig.\
\ref{Fig:psi1 psi2}, matches exactly the observed greater
stability of the island in the pseudoclassical phase space for
$T=60.5\mu$s. This is consistent with Peres's identification of
the behaviour of the fidelity as reflecting stability properties
of the phase space in the semiclassical limit \cite{peres93}, even
though our experiment (and numerics) operate in a pseudoclassical
regime which is far from semiclassical.

\begin{figure}
\begin{center}
\includegraphics[width=3.3in]{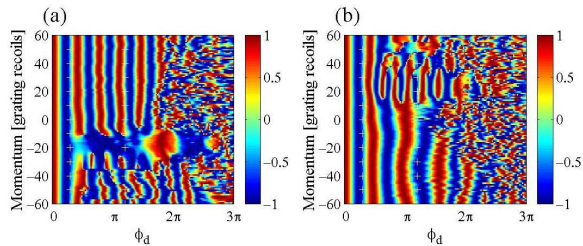}
\end{center}
\caption{Plots of $\cos(\phi_I + N\delta\phi_d)$ against
$\phi_{d}$ with $N=20$, for (a) $T = 60.5\mu$s and (b) $T =
74.5\mu$s. Crosses mark the boundaries of the experimental range.}
\label{Fig:zebra}
\end{figure}

The position of the islands in pseudoclassical phase space in
Fig.\ \ref{Fig:phasespace} indicates the region of the quantum
accelerator modes' spatial localization. For $T=60.5\mu$s this is
where there is zero standing light wave intensity, whereas when
$T=74.5\mu$s, it is where the intensity, and hence phase shift,
are maximal \cite{darcythesis}. We thus expect the modulation of
the $P_{b}$ interference term in Eq.\ (\ref{eq:simons
calculation}), $\cos(\phi_I+ N\delta\phi_d + \theta)$, to have a
strong dependence on $\phi_{d}$ for the momenta at which
accelerator modes are found when $T=74.5\mu$s, but not when
$T=60.5\mu$s. This is confirmed by Figs.\ \ref{Fig:zebra}(a) and
\ref{Fig:zebra}(b), where $\cos(\phi_I+ N\delta\phi_d+\theta)$  is
plotted as a function of $p$ and $\phi_d$ for constant $\theta$
(set to 0 for convenience) for $T=60.5\mu$s and $T=74.5\mu$s,
respectively. At the accelerator mode momentum, the value of
$\cos(\phi_I+ N\delta\phi_d)$ at $T=60.5\mu$s is almost
independent of $\phi_{d}$, whereas at $T=74.5\mu$s there is an
approximate frequency doubling, relative to other momenta. We can
now explain the presence or absence of interference fringes in
Fig.\ \ref{Fig:experiment theory} in terms of this difference in
$\phi_{d}$ dependence. As Eq.\ (\ref{eq:simons calculation})
determines the population for one specific value of $\phi_d$ only,
the observable momentum distribution is $\overline{P}_{b}(p)= \int
d \phi_{d} D(\phi_{d})P_{b}(\phi_{d},p)$, where $D(\phi_{d})$ is a
distribution describing the proportion of atoms experiencing a
particular $\phi_{d}$. Although for a single value of $\phi_{d}$
the visibility of the fringes at both $T=60.5\mu$s and
$T=74.5\mu$s is high,  Figs.\ \ref{Fig:zebra}(a) and
\ref{Fig:zebra}(b) show that integration over the experimental
range of $\phi_{d}$ causes a greater reduction in visibility at
$T=74.5\mu$s than at $T=60.5\mu$s. At large $\phi_d$ (greater than
occurs in our experiment), there is a breakdown of all structure
in plots of $\cos(\phi_I+ N\delta\phi_d)$. Comparing with Figs.\
\ref{Fig:psi1 psi2}(a) and \ref{Fig:psi1 psi2}(c), we observe this
to coincide with a falloff in $A$. Note, however, that as $\phi_I$
is determined numerically from complex interference terms, one
should be careful about attaching significance to values of
$\cos(\phi_I+ N\delta\phi_d)$ for which the corresponding value of
$A$ is close to zero.

In summary, we have performed a Ramsey-type interference
experiment and thus demonstrated the coherence of the production
of quantum accelerator modes, and hence their suitability for
applications in atom interferometry. Numerically, we have found
the accelerator modes to correspond to regions of greater quantum
stability, as quantified by the fidelity. This is consistent with
the presence of stable regions in the phase space of a recently
proposed pseudoclassical limit of $\delta$-kicked accelerator
dynamics, rather than the globally chaotic behaviour of the
semiclassical limit. These regions dictate the position of the
accelerator modes' spatial localization, allowing us to explain
the lack of fringes for the accelerator mode at certain pulse
periods, due to the experimental range of kicking strengths. Our
investigation of coherence in quantum accelerator modes has
allowed observation of their quantum-stable dynamics in this
classically chaotic system.

We thank R. Bach, K. Burnett, S. Fishman, I. Guarneri,  L.
Rebuzzini, and S. Wimberger for stimulating discussions. We
acknowledge support from the UK EPSRC, the Paul Instrument Fund of
The Royal Society, the EU through the TMR `Cold Quantum Gases'
network (contract no.\ HPRN-CT-2000-00125) and the Marie Curie
fellowship program (D.C.), the DAAD (S.S.), the Royal Commission
for the Exhibition of 1851 (M.B.d'A.), and Christ Church, Oxford
(R.M.G.)


\begin{thebibliography}{99}
\bibitem{amcomment}
Here $F$ is the total electronic plus nuclear angular momentum,
and $m_{F}$ is the projection on the quantization axis.
\bibitem{phinote}
As $\phi_{d}^{a}$ is determined by $\phi_{d}$, $\hat{F}_{a}$ is
written as a function of $\phi_{d}$.
\bibitem{visibility} We define the visibility by
$V = (S_{\mbox{\scriptsize max}} - S_{\mbox{\scriptsize
min}})/(S_{\mbox{\scriptsize max}} + S_{\mbox{\scriptsize min}})$,
where $S$ is the accelerator mode population in state $|b\rangle$.
\bibitem{stability}
Note that
$\delta\phi_{d}=\phi_{d}(1-\delta_{L}^{b}/\delta_{L}^{a})$ grows
linearly with $\phi_{d}$. Thus, as $\phi_{d}$ increases, so does
the difference between $\hat{H}_{a}$ and $\hat{H}_{b}$.
\end{thebibliography}
\end{document}